\documentclass[letterpaper, 10 pt, conference]{ieeeconf}

\IEEEoverridecommandlockouts                              % This command is only needed if
\usepackage{graphics} % for pdf, bitmapped graphics files
\usepackage{epsfig} % for postscript graphics files
\usepackage{mathptmx} % assumes new font selection scheme installed
\usepackage{times} % assumes new font selection scheme installed
\usepackage{amsmath} % assumes amsmath package installed
\usepackage{amssymb}  % assumes amsmath package installed
\usepackage{bm}
\usepackage{flushend}
\usepackage{stfloats}
\usepackage{fancyhdr}

\title{\LARGE \bf
Development of an Autonomous Sanding Robot with Structured-Light
Technology}
	
\author{Yingxin Huo, Diancheng Chen, Xiang Li, Peng Li, and Yun-Hui Liu % <-this % stops a space
\thanks{This work was supported by the CUHK T Stone Robotics Institute and the Innovation and Technology Commission of Hong Kong under Grant no.UIM/335 and in part by the VC Fund 4930745 of the CUHK T Stone Robotics Institute.(Corresponding author: Xiang Li (xiangli@cuhk.edu.hk), Co-corresponding author: Peng Li (peng.li@hit.edu.cn)) }% <-this % stops a space
\thanks{Y. Huo, D. Chen, X. Li and Y.-H. Liu are with the Department of Mechanical and Automation Engineering, The Chinese University of Hong Kong, Hong Kong S.A.R.  X. Li is also with the CUHK Shenzhen Research Institute (SZRI), China.  }
\thanks{P. Li is with the CUHK SZRI and the Department of Mechanical and Engineering and Automation, Harbin Institute of Technology (Shenzhen).}
}

\begin{document}

\maketitle
\thispagestyle{fancy} 
\pagestyle{fancy} 
\chead{\Large CONFIDENTIAL. Limited circulation. For review only.}

\headsep 10pt   
\cfoot{\large Manuscript 660 submitted to 2019 IEEE/RSJ International Conference
	\\ on Intelligent Robots and Systems (IROS). Received March 1, 2019.}
\footskip 30pt
\voffset = -20pt
\renewcommand{\headrulewidth}{0pt}

%%%%%%%%%%%%%%%%%%%%%%%%%%%%%%%%%%%%%%%%%%%%%%%%%%%%%%%%%%%%%%%%%%%%%%%%%%%%%%%%
\begin{abstract}
%Under the circumstance of high labor price and people increasingly care about workers' health.
%The need to use industry robot to do sanding operation is obvious.
Large demand for robotics and automation has been reflected in the
sanding works, as current manual operations are labor-intensive,
without consistent quality, and also subject to safety and health
issues.
While several machines have been developed to automate one or two
steps in the sanding works, the autonomous capability of existing
solutions is relatively low, and the human assistance or supervision
is still heavily required in the calibration of target objects or
the planning of robot motion and tasks.
%There are three basic issues in
%autonomous sanding robot, %1) can not complete the task without the
%CAD model of the sanded object. 2) can not sand the object without
%setting the sanding trajectory before sanding. 3) one sanding robot
%can only sand one type of object with simmilar profile.
%Based on
%these deficiencies of current sanding robot, this paper present a
%fully autonomous sanding robot which could sand all faces of a
%general object without human assistance.
This paper presents the development of an autonomous sanding robot,
which is able to perform the sanding works on an unknown object
automatically, without any prior calibration or human intervention.
The developed robot works as follows. First, % With structured-light
%camera, the profile of any sanded object can be scanned and modeled.
the target object is scanned then modeled with the structured-light
camera. Second, the robot motion is planned to cover all the
surfaces of the object with an optimized transition sequence. Third,
the robot is controlled to perform the sanding on the object under
the desired impedance model.
%Based on the point cloud model, motion planning algorthm is proposed
%for the sanding robot so that the robot can sand all faces
%automatically.
%Also, with the modular design of the structure of sanding robot, the
%holder could be changed and the whole system could be used for any
%kind of object.
%During the sanding procedure, for testing the sanding quality,
%structured-light camera is applied for quality detection after
%sanding.
A prototype of the sanding robot is fabricated and its performance
is validated in the task of sanding a batch of wooden boxes. With
sufficient degrees of freedom (DOFs) and the module design for the 
end effector, the developed robot is able to provide a general
solution to the autonomous sanding on many other different objects.
%The experiment results are given out to verity the
%autonomous robot scheme. The experiment result shows that the
%proposed autonomous sanding robot could recognize the sanded object
%and plan the motion according to the processed point cloud data
%(PCD) and then sand all faces of the object automatically. The
%dynamic stability of the system is proved with Lyapunov methods
%precisely.
\end{abstract}

%%%%%%%%%%%%%%%%%%%%%%%%%%%%%%%%%%%%%%%%%%%%%%%%%%%%%%%%%%%%%%%%%%%%%%%%%%%%%%%%
\section{INTRODUCTION}
\begin{figure}
    \centering
    \begin{minipage}[c]{0.5\textwidth}
        \centering
        \includegraphics[height=4.3cm,width=8.3cm]{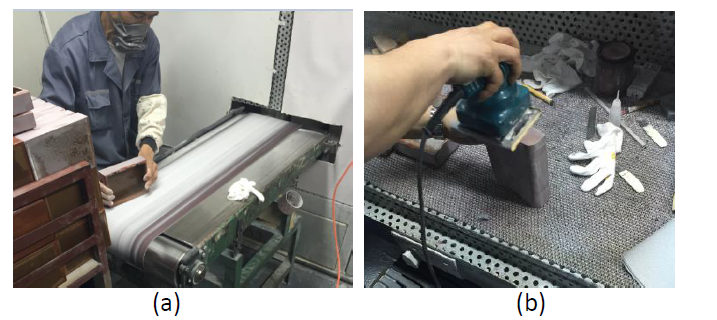}
    \end{minipage}%
    \caption{Manual sanding operations (a) Sanding the wooden box on a sand paper machine (b) Sanding the wooden box with a portable tool}
    \label{manualWorks}\vspace{-0.8cm}
\end{figure}

%Sanding is a traditional technology traced back to hundreds of years
%ago.
Sanding is a common operation in the manufacturing industry, where
workers stand next to a spinning sanding belt or use a portable tool
to remove the coating from a target object (see Fig.
\ref{manualWorks}). Current sanding operations are mainly performed
by human workers, and disadvantages include: i) the
labor-intensive nature, ii) the inconsistency in quality, iii) the
unhealthy working environment (e.g., dust, noise) and safety issues
(e.g., fatigue, injury). Therefore, the manual sanding is now with
the serious problems about the shortage of skilled workers and their rapidly increasing wages.

%- set Fig 1 bigger -
%- use vspace command to reduce the space -,
%The harmful effect it takes to workers cannot be eliminated until the development of industry robot.
To address the problems associated with the manual sanding. Some
industry polishing machine for a specific application was developed
since 2000 [1]-[3] by Nagata. In the beginning, the polishing work
relied on calibration before the process, then with the CAD model
of the sanded objects, the sanding trajectory could be given out
according to the CAD model, but still, CAD model should be set and the path should be calculated in advanced. With the research into control theory, in  [2]
feedforward controller based on the superior calculated
trajectory was proposed . It was the first time the controller was
adopted in generating the trajectory though it was based on the
calibration data collected before the experiment. In [3] automatic polishing system was presented, the polishing path was generated through file reading or robot teaching and then drew the polishing path on the mold surface, the force controllable robot was used to polish the face.
After trying the particular shape
model, free-formed surface model was
considered [4], but the CAD model of the sanded object was necessary and the
method could only be used for one curved plane. In all these applications, human assists [1]-[5] were indispensable, even if
with the CAD/CAM system, the system could only generate the sanding
trajectory for one flat or curved plane based on the data collected in advanced.  %- delete *paper* %

%- check impedance control for sanding -
As the robot is controlled to interact with the target surfaces of the sanded object, the interaction force should be regulated to guarantee the quality of sanding. There were several control methods proposed to solve the problem of interaction force and position control in the past two to three decades. Before automating the process of sanding, the automatic process using impedance controller, such as deburring, which is less stress on the compliant interaction force was proposed [6][7]. In sanding and polishing operations, there are three classes of the method to cope with the interaction force between the sanding end effector and the surface. When there is no force sensor available, the reaction force observer (RFOB) could be designed into the PD control loop to estimate the force [8].  The second method is to try to decouple the force and position control with controller design and mechanical design [9-11]. The third method is to use the impedance model to build up the impedance controller [12].
Impedance controller [12][13] was adopted to control the interaction force in many traditional and modern industry robots and adaptive control are effective according to recent research in controller design [14]. In our application, because of the interaction force is applied at the end point of the robot, the force sensor is available, position and force control are coupled, so the adaptive impedance controller is adopted under the task space [15].

However, the autonomous capability of existing systems for automatic sanding is
relatively low, in the sense that a lot of
human involvement or assistance is still required in some specific
steps of sanding. For example, the manual calibration is usually required to obtain the CAD model of sanded objects beforehand, or humans should specify the sequence of sanding on different surfaces of the object, or the desired trajectory of the robot should be manually set according to the human's experience, or the sanding quality is assessed by humans.

In this paper, a new autonomous sanding robot is proposed with the structured-light technology, which is to automate the labor-intensive works and hence systematically address the problems associated with manual operations. Compared with existing systems, the autonomous capability of the proposed robot is significantly improved in terms of the following aspects. First, the structured-light scanner is employed to scan the unknown object and construct its model with point clouds.
%1) With the structured light camera the initial status of the sanded object is identified and according to point cloud data the best motion is planned.
Second, the model-based planner is proposed to specify the reference motion of the end effector such that it covers all the surfaces of the object under an optimized sequence.
%2) An adaptive impedance controller is developed to regulate the interaction force between the sanded object and the sandpaper in real-time.
Third, the adaptive impedance controller is developed to achieve the desired impedance model and hence drive the robot to carry out the sanding task.
%3) A time-varying desired position is specified according to the amount of feed rate in the desired impedance model, so as to maintain the contact between the soft coating layer and the sandpaper also avoid physical damages to the sanded object.
Fourth, the quality evaluation is performed by assessing the changes of three-dimensional information of the sanded surface.
%4) Also, the quality evaluation method is given out with the structure light camera.
Note that no human assistance or supervision is required in the above steps or the transition between steps. The stability of the closed-loop system is rigorously analyzed with Lyapunov methods, such that the proposed adaptive impedance controller is theoretically grounded. A prototype of the proposed robot is fabricated and its performance is validated in the task of sanding a batch of wooden boxes.
%Firstly, the manual sanding operations for the small and delicate object are introduced.
%Fig. 1(a) shows that the worker is polishing the wooden box on a sandpaper machine,
%where the worker pushes the box against the sand paper in order to grind the priming paint.
%Next, the worker uses a portable polishing tool to polish surface of the wooden box to remove the reflective layer on it,
%which could be seen in  Fig. 1(b).
%
%
%To complete the task, the prototype of sanding robot is designed as shown in Fig. 2.
%In this paper, with the structured light camera the point cloud model could be generated.
%Then, to make industrial robot complete tasks efficiently, two kinds of planner should be considered:
%single-query motion planner and multi-query motion planner. In this paper,
%we define the single-query motion planner as the combination of path planner
%which generate configuration nodes that robot will pass by and trajectory planner which design proper speed and acceleration
%for every joint of the robot. And the multi-query planner designs the execution sequence for multi-tasks.

\section{BACKGROUND}
\subsection{Robot Kinematics and Dynamics}
The forward kinematic model of a robot manipulator can be given as
\begin{eqnarray}
&\bm x=\bm h (\bm q),
\end{eqnarray}
where $\hspace{-0.05cm} \bm x\hspace{-0.05cm}\in{\Re}^m\hspace{-0.05cm}$ represents the position of the robot end effector in task space (e.g. Cartesian space or vision space), $\hspace{-0.05cm}\bm q\in {\Re}^n\hspace{-0.05cm}$ is the vector of robot joint angles, and $\bm h(\cdot)$ denote the nonlinear functions. Then, the velocity of the robot end effector in task space is related to the joint-space velocity of the robot as
\begin{eqnarray}
&\dot{\bm x}=\bm J (\bm q)\dot{\bm q},
\end{eqnarray}
where $\bm J (\bm q)\in{\Re}^{m\times n}$ is the Jacobian matrix from joint space to task space.

The dynamic model of a robot manipulator can be described as
\begin{equation}
\bm{M}(\bm{q})\bm{\mathop q\limits^{..}}  + \bm{C}(\bm{q},\bm{\mathop q\limits^. })\bm{\mathop q\limits^.}  + \bm{g}(\bm{q})  =\bm u+ \bm \tau_e, \label{dynamic model}
\end{equation}
where $\bm{M}(\bm{q})\bm{\mathop q\limits^{..}}  \in {\Re^n}\ $, $\bm{C}(\bm{q},\bm{\mathop q\limits^. })\bm{\mathop q\limits^.} \in {\Re^n}\ $, and $\bm{g}(\bm{q}) \in {\Re^n}\ $ denote the inertia torque, the centripetal and coriolis torque, and the gravitational torque of the robot respectively, $\bm u$ denotes the control input torque, and $\bm\tau_e $ denotes the interaction torque. Note that $\bm\tau_e=\bm J^T(\bm q)\bm f_e$, and $\bm f_e\in\Re^m$ is the interaction force between the robot end effector and the environment.

Some important properties of the dynamic model (\ref{dynamic model}) are listed as follows.

(i) The dynamic parameters $\bm M(\bm q)$,  $\bm C(\bm q,\dot{\bm q})$ and $\bm g(\bm q)$ are bounded;

(ii) The matrix $\bm M(\bm q)$ is symmetric and positive definite;

(iii) The matrix $\dot{\bm M}(\bm q)\hspace{-0.05cm}-\hspace{-0.05cm}2\bm{C}(\bm{q},\dot{\bm{q}})$ is skew-symmetric.

%(iii) The first three terms of (\ref{eq}) are linear in a set of physical parameters $\bm\theta\hspace{-0.05cm}=\hspace{-0.05cm}\left [\bm\theta_1, \cdots, \bm\theta_k\right ]^T\hspace{-0.1cm}\in\hspace{-0.1cm}\bm\Re^k$ as:
%\begin{eqnarray}
%&\bm M(\bm q)\ddot{{\bm q}}\hspace{-0.05cm}+\hspace{-0.05cm}\bm
%C(\bm q, \dot{\bm q})\dot{\bm q}\hspace{-0.05cm}+\hspace{-0.05cm}\bm
%g(\bm q)=\hspace{-0.05cm}\bm{Y}(\bm{q},\dot{\bm{q}},\dot{\bm{q}},\ddot{\bm{q}})\hspace{-0.05cm}\bm\theta,
%\end{eqnarray}
%where $\bm{Y}(\cdot)\hspace{-0.1cm}\in\hspace{-0.1cm}\Re^{n\times k}$ is a known dynamic regressor matrix.

In this paper, the dynamic parameters, i.e. $\bm M(\cdot)$ , $\bm C(\cdot)$, and $\bm g(\cdot)$ are assumed to be unknown and will be estimated by on-line adaptation laws.

\subsection{Impedance Control}
In impedance control, the control objective is specified as a desired impedance model in task space as
\begin{eqnarray}
&\bm M_d(\ddot{\bm x}-\ddot{\bm x}_d)\hspace{-0.05cm}+\hspace{-0.05cm}\bm
C_d(\dot{\bm x}-\dot{\bm x}_d)\hspace{-0.05cm}+\hspace{-0.05cm}\bm K_d(\bm x-\bm x_d) =\hspace{-0.05cm}\bm f_e\hspace{-0.05cm}-\hspace{-0.05cm}\bm f_d,\label{TargetImpedance}
\end{eqnarray}
where $\bm x_d\hspace{-0.05cm}\in\hspace{-0.05cm}\Re^m$ denotes the time-varying desired trajectory in task space, $\bm f_d\hspace{-0.05cm}\in\hspace{-0.05cm}\Re^m$ is the desired force, $\bm M_d, \bm C_d, \bm K_d\in\Re^{m\times m}$
denote the desired inertia, the desired damping, and the desired stiffness
matrices respectively, which are diagonal and constant. The desired impedance model (\ref{TargetImpedance}) describes a dynamic relationship between the position error and the interaction force.

The desired impedance model (\ref{TargetImpedance}) can be rewritten as
\begin{eqnarray}
&\Delta\ddot{\bm x}\hspace{-0.05cm}+\hspace{-0.05cm}\bm M_d^{-1}\bm
C_d\Delta\dot{\bm x}\hspace{-0.05cm}+\hspace{-0.05cm}\bm M_d^{-1}\bm K_d\Delta\bm x-\bm M_d^{-1}\Delta\bm f=\hspace{-0.05cm}\bm 0, \label{TargetImpedance2}
\end{eqnarray}
where $\Delta\bm x=\bm x-\bm x_d$ and $\Delta\bm f=\bm f_e\hspace{-0.05cm}-\hspace{-0.05cm}\bm f_d$.
%By introducing the position error $\bm \Delta \bm{r=r-r_{d}}$,

Next, by introducing the matrices $\bm\Lambda$, $\bm\Gamma$, and the vector $\Delta \bm f_l$ as [16]
\begin{eqnarray}
&\bm\Lambda +\bm\Gamma = \bm M_d^{-1}\bm C_d,\\ \label{Lambda1}
&\bm\Lambda \bm\Gamma =\bm M_d^{-1}\bm K_d,\\ \label{Lambda2}
&\Delta\dot{\bm f}_l+\bm\Gamma\Delta \bm f_l= \bm M_d^{-1}\Delta\bm f, \label{Lambda3}
\end{eqnarray}
an impedance vector can be proposed as
%we can obtain an augmented vector as $\bar{\bm w}=\bm\Delta \dot{\bm r}+\bm\Lambda \bm\Delta \bm{r}-\dot {\bm f}_{\bm l}-\bm\Gamma {\bm f}_{\bm l}=\dot {\bm z}+\bm\Gamma {\bm z}$, where
\begin{eqnarray}
&\bm z=\Delta\dot{\bm x}+\bm\Lambda\Delta\bm x-\Delta \bm f_l,\label{vectorz}
\end{eqnarray}
such that the left side of (\ref{TargetImpedance2}) equals to
\begin{eqnarray}
&\Delta\ddot{\bm x}\hspace{-0.05cm}+\hspace{-0.05cm}\bm M_d^{-1}\bm
C_d\Delta\dot{\bm x}\hspace{-0.05cm}+\hspace{-0.05cm}\bm M_d^{-1}\bm K_d\Delta\bm x-\bm M_d^{-1}\Delta\bm f
=\dot{\bm z}+\bm\Gamma\bm z.\label{temp1}
\end{eqnarray}

From (\ref{temp1}), the impedance vector $\bm z$ can be treated as the low-pass-filtered signal of the desired impedance model. In this paper, the control objective is formulated as
$\bm z\rightarrow\bm 0$ as $t\rightarrow\infty$, indicating the realization of the desired impedance model in the low-frequency range.

\subsection{Structured-Light Technology}
\begin{figure}[!h]\vspace{-0.3cm}
	\centering
	\begin{minipage}[c]{0.5\textwidth}
		\centering
		\includegraphics[height=5.5cm,width=9.7cm]{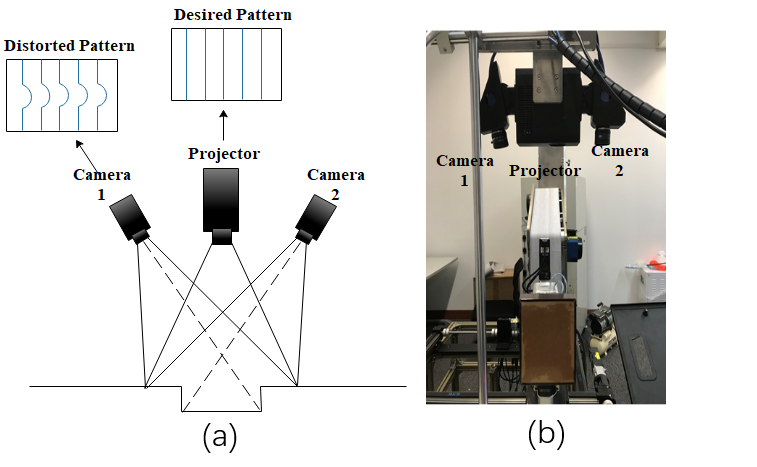}
	\end{minipage}%
	\caption{(a) Working principle of a structured-light system; (b) A structured-light scanner.}
    \label{structured}\vspace{-0.2cm}
\end{figure}

A collection of projected rays in the spatial direction is called structured light. The structured-light system is characterized by a set of projectors and cameras. The working principle of the structured-light system is shown in Fig. \ref{structured}(a).
The projector projects a specific patterned image onto the target object first, and the cameras capture the distorted image due to three-dimensionally shaped surface, then the processor processes the raw data from the camera and outputs the reconstructed position of the object.  For the sake of avoiding obstructions, two cameras are usually installed to capture the image from different angles of view. The advantages of structured-light technology are listed as follows: firstly,  since the structured light actively projects the coded light, it is very suitable for using in scenes with insufficient illumination (even no light) and lack of texture(such as sanded object in this paper). Secondly, the structured light projection pattern is specially designed to achieve high measurement accuracy within a certain range.
%The grating in the projector shown in Fig.2(a) is changed after captured by the camera because of the depth change.
%The fundamental of three-dimensional structured light is to obtain the three-dimensional structure of the object by optical means, and then use this information for a further application. Three-dimensional outline of the sanded object could be generated by the scanning of the structured-light 3D camera.
% label each figure
% - set Fig. 2 bigger -

A structured-light scanner is shown in Fig. 2(b).
The scanned data is recorded in the form of dots, named point clouds, each of which contains three-dimensional coordinates, and some may contain color information (RGB) or reflection intensity information (intensity). The applications of the structured-light scanner include motion tracking [17], freeform surfaces inspection [18] and anthropometric parameter estimation[19] and so on. With advance of computer vision and electronic devices, structured-light scanner become more popular in industry robots.

\section{Autonomous Sanding Robot}
\begin{figure}[!h]\vspace{-0.3cm}
	\centering
	\begin{minipage}[c]{0.5\textwidth}
		\centering
		\includegraphics[height=7.5cm,width=8.3cm]{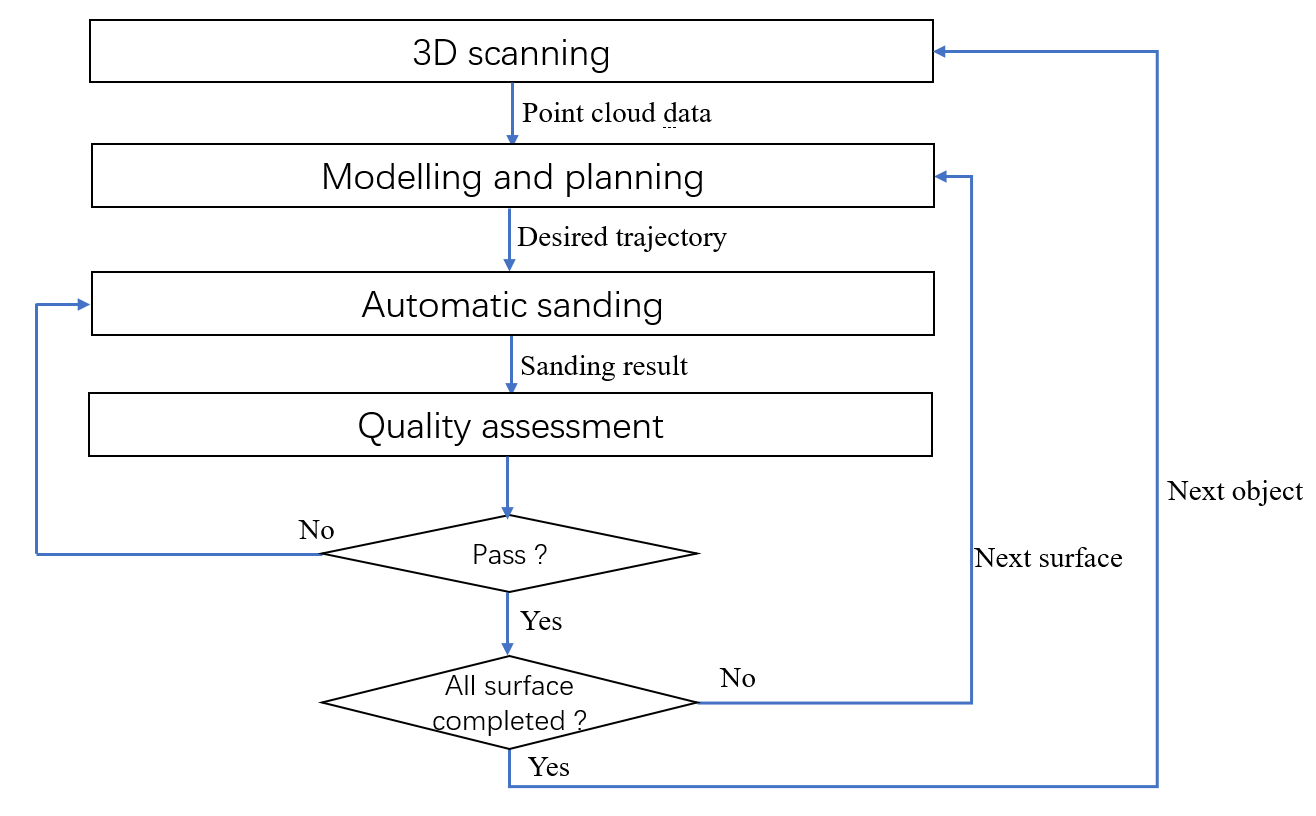}
	\end{minipage}%
	\caption{Overall structure of the autonomous sanding robot}\label{overall}
	\vspace{-0.6cm}
\end{figure}

In this paper, a robotic system is developed to achieve the autonomous sanding with the work flow shown in Fig. \ref{overall}. First, %based on structured-light technology the point cloud data can be generated and then the data is sent to the computer to complete the modeling task.
a structured-light scanner is used to scan the target object and reconstruct its CAD model with the point clouds. Second, a planning method is proposed to generate the sanding sequence for different surfaces of the object by referring to the CAD model. Third, an adaptive impedance controller is developed to drive the robot to perform the sanding task by following the optimized sequence. At the end of each sanding procedure,
%According to the model, the path is planned to cover all the surfaces of the object with an optimized transition sequence.
%During the sanding procedure, adaptive impedance controller is adopted.
the sanded surface is automatically assessed by employing the structured-light technology.
%After finishing the sanding of all faces, the quality assessing based on the PCDs before and after sanding can decide whether go bank to sanding procedure.
If the sanding quality passes the assessment, the robot proceeds to sand the next surface or the next object. Otherwise, the same surface is sent back for re-sanding till its quality meets the requirement.
%If the sanded object does not pass the test, the object needs to be sanded again. If it passes the qualification, ether amount the same sanded object then the program goes back to the planning procedure, or mount another object then the program goes back to the scanning procedure.
%%If the all surfaces completed,

\subsection{Mechanical Design}
\begin{figure}[!h]
	\centering\vspace{-0.2cm}
	\begin{minipage}[c]{0.5\textwidth}
		\centering
		\includegraphics[height=5.5cm,width=8.3cm]{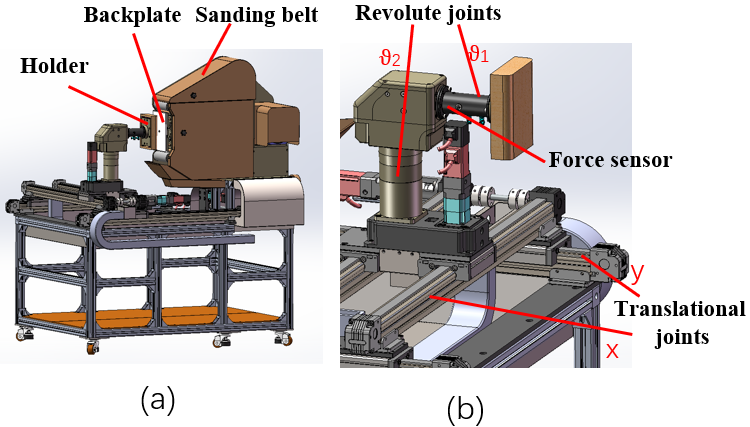}
	\end{minipage}\vspace{-0.3cm}
	\caption{The CAD model of the autonomous sanding robot (a) The overall structure; (b) The robot arm and the sanding end effector.}
	\label{CADmodel}\vspace{-0.2cm}
\end{figure}

The mechanical structure of the sanding robot is shown in Fig. \ref{CADmodel}(a), consisting of the robot arm, the end effector, the mechanism of sanding belt, and the sensing and control module. To achieve the sanding task, the robot is controlled to grasp the object then manipulate it to physically interact with the sanding belt. The robotic arm has four degrees of freedom (DOFs), that is, two translational joints (to position the object) and two revolute joints (to adjust its orientation with respect to the sanding belt). The ATI force sensor is mounted on the robot arm shown in Fig. \ref{CADmodel}(b) to collect interaction force between backplate and sanded object during the operation. The control module consists of four servo motors to control the four joints shown in Fig. \ref{CADmodel}(b), IO module and frequency conversion device. IO module is implemented to send switching signals to pneumatic components and frequency conversion device is used to control the speed of sanding. %% - introduce the sensing and control module -

As seen in Fig. \ref{CADmodel}, an air cylinder push-pull mechanism is designed and installed on the end effector to grasp the sanded object. In particular, the piston of the cylinder can push a linkage mechanism that drives the four rubber fingers simultaneously to grasp the object. When the robotic arm comes back to the docking position, the piston pulls the linkage mechanism reversely and hence the grasping is released. An adjustable mechanism is also designed for the back plate which is used to make sure the backplate is perpendicular to the ground. The backplate and holder are controlled by pneumatic drivers. The back plate can make sure that the object and the sanding belt contact tightly with each other.
%This design is based on modular, if the sanded object changed, the holder could be replaced to adapt to the sanded object.  The back plate of the robot can realize the compliant contact between the sanding paper and sanded object in structure level.
Note that such design in Fig. \ref{CADmodel} is general and can be easily extended to many different sanded objects. Because i) it has four DOFs which is sufficient for the sanding of objects with regular shapes, and ii)
the end effector can be customized and replaced to suit the specific size and shape of objects.

%During the sanding process, the most important is the control of sanding force and feed amount, i.e., position. There are two different strategies in control of sanding or polishing process, active compliance control [21,22,23], which is focused on the active force control, or passive compliance control [24,25], which is focused on the natural compliance of mechanical design. In our application, both active and passive compliance are considered. The adaptive impedance controller is designed to ensure the proper force and position in the sanding process. A passive compliance backplate is designed to ensure tight contact in the sanding process.

%\subsection{Structured Light Camera}
\subsection{Modeling and Planning}
\begin{figure}[!h]
	\centering\vspace{-0.7cm}
	\begin{minipage}[c]{0.5\textwidth}
		\centering
		\includegraphics[height=3.5cm,width=8.5cm]{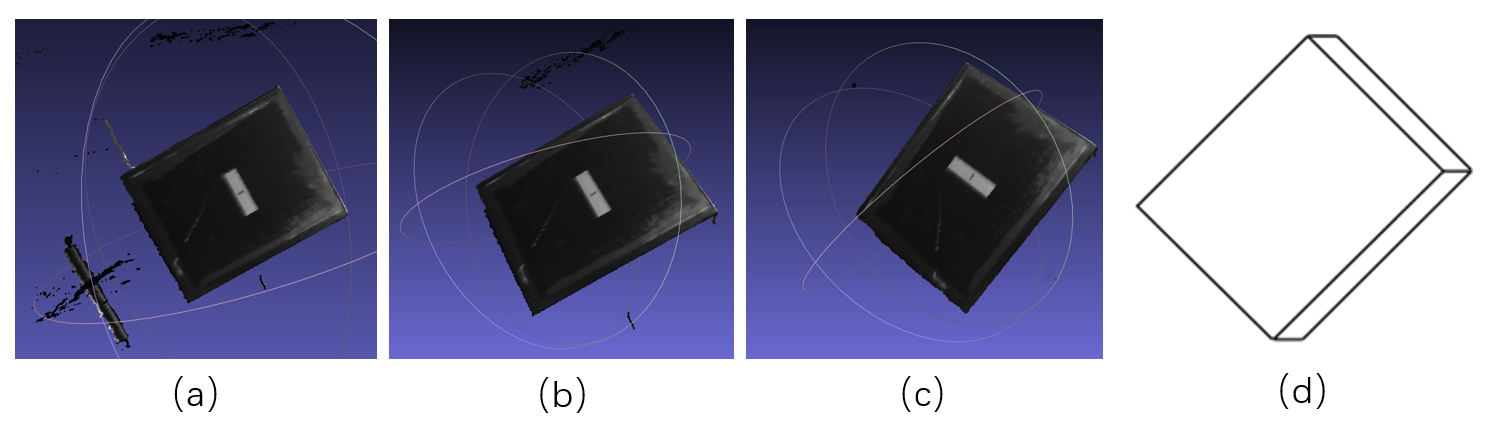}
	\end{minipage}
	\caption{Modeling with the structured-light scanner (a) Raw data; (b) After the processing with the field limits filter; (c) After the processing with the SOR Filter; (d) CAD model}
	\label{pointcloud}
\end{figure}

%Because the sanded objects always have more than one face and with one structured light camera could only detect a small part of the sanded object, also because the original point cloud data (PCD) collected always contain noises, so
%point cloud noise filtering, registration and splicing are necessary steps in modeling the sanded object.
The proposed robot is able to sand the object without knowing the CAD model beforehand. The unknown object is scanned by the structured-light scanner first, then its CAD model is constructed with the point clouds generated by the scanner. In particular, the modeling consists of three steps (see Fig. \ref{pointcloud}).  %In this application, the data collected by the structured-light camera is point cloud data, so Point Cloud Library (PCL) is used to cope with the data.
First, the field limits filter is set up to ensure the data captured are all from the sanded object but not from the sanding robot holder. Second, the Statistical Outlier Removal(SOR) filter [20] is applied to remove outliers in the collected data. Third, the 3D registration method [21-23] is employed to retrieve the whole structure of the sanded object. %The point cloud data during these three steps are shown in Fig. \ref{pointcloud}. Fig. \ref{pointcloud}(a) is the raw data collected from the 3D scanner,  Fig. \ref{pointcloud}(b) is the point clouds after field limits filter and Fig. \ref{pointcloud}(c) is the data after the SOR filter. Finally, Fig. \ref{pointcloud}(d) is the CAD model after registration.
\vspace{-0.3cm}

The basic idea in registration is to convert every data in different coordinate systems into one unified coordinate system and minimize the error, that is, %. The calculation of error as equation (11):
%\begin{equation}
$\min\{\sum\limits_{i = 1}^k{||\bm y_{ai} - \bm R(\bm y_{oi}) - \bm{T}|{|^2}}\}$, 
%\end{equation}
where $ \bm{R} $ denotes the rotation matrix, $ \bm{T} $ denotes the translation vector, $\bm y_{oi}$ and $\bm y_{ai}$ represent the points under the original  coordinate system and the  points under aimed coordinate system respectively. The mapping between two points is represented as $\bm y_{ai}=\bm R(\bm y_{oi}) +\bm{ T }$. Therefore, given $k$ pairs of matching points, it is to seek the rotation matrix $ \bm{R} $ and translation vector $ \bm{T} $  to minimize the error. In this paper, in order to scan different faces of the target object, the object is manipulated to rotate around $ \theta _2\ $ (see Fig. \ref{CADmodel}(b)) then the rotation matrix $ \bm{R} $ and the translation vector $ \bm{T} $ are computed with the algorithm of Iterative Closest Point (ICP) [22] under such configuration. %Finally, splice then to get the entire profile of sanded object as (12) show.
%- see if a function can be used to represent the mapping from the collection of point clouds $\bm x_p$ to the function of CAD model $\bm y$. -
%\begin{eqnarray}
%CAD\_MODEL = PCD\_FACE1 + PCD\_FACE2({\bm{R}_1},{\bm{T}_1})\nonumber \\+ PCD\_FACE3({\bm{R}_2},{\bm{T}_2}) + ......... + PCD\_FACEn({\bm{R}_n},{\bm{T}_n})
%\end{eqnarray}

With the constructed CAD model, the model-based planner is proposed to generate the desired trajectory for the robot. The planner is designed with the hierarchical structure. In the upper level, the multi-query method is used to optimize the sanding sequence for all the surfaces of the object. In the lower level, the single-query method is used to optimize the motion of the robot from one configuration to another. 

In the single-query method, the objective is to generate the collision-free path with the desired speed and the acceleration. To obtain the collision-free path, consider the vertices of polyhedrons which enclose target object and platform CAD model: ${\bm V_o} = \{ {v_{{o_1}}},{v_{{o_2}}}...\} $,
${\bm V_p} = \{ {v_{{p_1}}},{v_{{p_2}}}...\} $, with their corresponding connect sequence: ${\bm S_o}$, ${\bm S_p}$ , and initial and goal configurations of the robot:
${\bm q_{start}}$,
${\bm q_{goal}}$. Then, the collision-free path of the robot is generated as
\begin{eqnarray}
\bm c = {F_p}({\bm V_o},{\bm V_p},{\bm S_o},{\bm S_p},{\bm q_{start}},{\bm q_{goal}}), \label{p1}
\end{eqnarray}
where
$ \bm c:[{\rm{0}},m] \to \bm Q$,
$ \bm c(0) = {\bm q_{start}}$,
$ \bm c(m) = {\bm q_{goal}}$,
$ \bm c(s) \in {\bm Q_{free}}$,
$\forall s \in [0,m]$,
${F_p}$ denotes the process of path planning,
$\bm Q$  denotes the configuration space,
${ \bm Q_{free}}$ denotes the configuration space which ensures collision-free criterion, s is the time step, and m is the length of time steps. The main idea of the method (11) is to connect the initial configuration and the goal configuration with a straight line then check if it is a collision-free path. 
The path of the robot should be updated if there exists a collision. In this paper, the collision checking is based on GJK algorithm, which iteratively generates simplex to determine whether the Minkowski difference of two convex shape includes the origin, and many algorithms are developed to deal with  collision avoidance, for complicated environment, Probabilistic Road Map (PRM), Rapidly-Exploring Random Tree(RRT) can be adopted, while for simple obstacles, specific rules can be set to guide the robot away from obstacles [\ref{planner3}]. After the collision-free motion is obtained, the method of Linear Segment with Parabolic Blend(LSPB) [\ref{planner2}] is used to further shape it to achieve the desired speed and the desired acceleration such that the speed of motor can be maintained in its maximum level for a certain duration (that is, with higher efficiency).  

In the multi-query method, the summation of weighted squared displacement is set as the objective function, as shown in (12), where $ \bm w $ is the weight vector to rescale the rotational distance and translational distance, h the index of sub-objective. Then, the Genetic algorithm is employed to obtain the sanding sequence with the minimum travel distance.
\begin{eqnarray}
{F_{cost}} = {\sum\limits_h^{} {\sum\limits_s^{} {\left\| { \bm w\left( {{\bm c_h}\left( {s + 1} \right) - {\bm c_h}\left( s \right)} \right)} \right\|} } ^2},\label{p3}
\end{eqnarray}

\subsection{Adaptive Impedance Control}\label{AA}
The control objective of the sanding robot is to realize the desired impedance model (\ref{TargetImpedance}) and hence carry out the sanding task on the target surface. First, a new
impedance vector is formulated in joint space as
\begin{eqnarray}
&\bm z_q=\dot{\bm q}-\dot{\bm q}_r
=\dot{\bm q}-\bm J^{+}(\bm q)(\dot{\bm x}_d-\bm\Lambda \Delta\bm x+\Delta \bm f_l),\label{temp2}
\end{eqnarray}
where $\bm J^{+}(\bm q)$ is the pseudo-inverse matrix, and
\begin{eqnarray}
&\dot{\bm q}_r=\bm J^{+}(\bm q)(\dot{\bm x}_d-\bm\Lambda \Delta\bm x+\Delta \bm f_l),\label{temp5}
\end{eqnarray}
is a reference vector.

From (\ref{vectorz}) and (\ref{temp2}), it can be obtained that
\begin{eqnarray}
&\bm z=\bm J(\bm q)\bm z_q.\label{temp4}
\end{eqnarray}
Hence, the convergence of $\bm z_q\rightarrow\bm 0$ implies that $\bm z\rightarrow\bm 0$, that is, the realization of the desired impedance model.

Using (\ref{temp2}), the dynamic model (\ref{dynamic model}) can be written as
\begin{eqnarray}
&\bm{M}(\bm{q})\hspace{-0.05cm}\dot{\bm{z}}_q\hspace{-0.05cm}+\hspace{-0.05cm}\bm{C}\hspace{-0.05cm}(\bm{q}, \dot{\bm q})\bm{z}_q\hspace{-0.05cm}+\hspace{-0.05cm}\bm{M}(\bm{q})\ddot{\bm
q}_r\hspace{-0.05cm}+\hspace{-0.05cm}\bm C(\bm q, \dot{\bm
q})\dot{\bm
q}_r+\bm g(\bm q)\nonumber\\
&=\bm \tau_e+\bm u.\label{temp3}
\end{eqnarray}
%\hspace{-0.05cm}+\hspace{-0.05cm}\bm{Y}_{\bm{q}}(\bm{q},\dot{\bm{q}},\dot{\bm{q}_{\bm{r}}},\dot{\bm{q}}_{\bm{r}})\hspace{-0.05cm}\hspace{-0.05cm}\bm{\psi} _{\bm q}=\hspace{-0.05cm}\bm\tau_{\bm e}\hspace{-0.05cm}+\hspace{-0.05cm}\bm \tau,
%\end{eqnarray}
%where  $\hspace{-0.05cm}\bm{Y}_{\bm{q}}(\bm{q},\dot{\bm{q}},\dot{\bm{q}_{\bm{r}}},\dot{\bm{q}}_{\bm{r}})\hspace{-0.05cm}\hspace{-0.05cm}\bm{\psi} _{\bm q}=\bm{M}(\bm{q})\hspace{-0.05cm}\ddot{\bm{q_{r}}}\hspace{-0.05cm}+\hspace{-0.05cm}\bm{C}\hspace{-0.05cm}\dot{\bm{q_{r}}}+\bm{g}(\bm{q})$, the vector $\bm{\psi} _{\bm q}\in \bm\Re ^{n\times 1}$ denotes the constant dynamic parameters, and the matrice $ \hspace{-0.05cm}\bm{Y}_{\bm{q}}\in \bm\Re ^{n\times n} $ is composed of joint angle, velocity and accelebration.

In this paper, it is assumed that the dynamic parameters $\bm{M}(\cdot)$, $\bm C(\cdot)$, and $\bm g(\cdot)$ are unknown. Then, the term $\bm{M}(\bm{q})\ddot{\bm
q}_r\hspace{-0.05cm}+\hspace{-0.05cm}\bm C(\bm q, \dot{\bm
q})\dot{\bm
q}_r+\bm g(\bm q)$ is approximated with the adaptive neural networks (NNs) techniques [\ref{NN}][\ref{NN1}]. The implementation
of the adaptive NNs eliminates the phase of offline training, and it
also eliminates the requirement of the exact structure of disturbances or robot dynamics. In particular, the radial-based-function (RBF) [\ref{RBF}], NNs is employed in this paper as
\begin{eqnarray}
&\bm M\hspace{-0.05cm}(\bm q)\ddot{\bm
q}_r\hspace{-0.05cm}+\hspace{-0.05cm}\bm C(\bm q, \dot{\bm
q})\dot{\bm
q}_r\hspace{-0.05cm}+\hspace{-0.05cm}\bm g(\bm q)
=\bm W\bm\theta(\bm q,\hspace{-0.05cm}\dot{\bm
q},\hspace{-0.05cm}\dot{\bm q}_r, \ddot{\bm
q}_r)\hspace{-0.05cm}+\hspace{-0.05cm}\bm E,\label{temp11}
\end{eqnarray}
where $\bm W$ is the ideal weight matrix, $\bm\theta(\cdot)$ is the activation function, and $\bm E$ is the
vector of approximation error.

Next, the adaptive impedance controller is proposed as
\begin{eqnarray}
&\bm u=-\bm K_z\bm z_q+\hat{\bm W}\bm\theta(\bm q, \dot{\bm q}, \dot{\bm q}_r, \ddot{\bm q}_r)-k_g\bm{sgn}(\bm z_q)-\bm \tau_e,\label{control}
\end{eqnarray}
where $\bm K_z\in \Re^{n\times n}$ is a diagonal and positive-definite matrix, $k_g$ is a positive constant, $\bm{sgn}(\cdot)$ is a sign function, and $\hat{\bm W}$ is the estimate of $\bm W$, which is updated by
\begin{eqnarray}
&\dot{{\hat{\bm W}}}_j^T=-{\bm L}_j\bm\theta(\bm q, \dot{\bm q}, \dot{\bm q}_r, \ddot{\bm q}_r)z_{qj},\label{NNupdate}
\end{eqnarray}
where $\hat{\bm W}_{j}$ represents the $j^{th}$ row vector of
$\hat{\bm W}$, $z_{qj}$ is the $j^{th}$ element of $\bm z_q$, and $\bm
L_{j}$ are diagonal positive definite matrices.

Substituting (\ref{control}) into (\ref{temp3}) and using (\ref{temp11}) yields
\begin{eqnarray}
&\bm M(\bm q)\dot{\bm
z}_q\hspace{-0.1cm}+\hspace{-0.1cm}\bm C(\bm q, \dot{\bm q})\bm
z_q\hspace{-0.1cm}+\hspace{-0.1cm}\bm K_z\bm
z_q\hspace{-0.05cm}+\hspace{-0.05cm}k_g\bm{sgn}(\bm z_q
)\nonumber\\
&\hspace{-0.0cm}+\Delta{\bm W}\bm\theta(\bm
q,\hspace{-0.05cm}\dot{\bm q},\dot{\bm q}_r, \ddot{\bm q}_r)\hspace{-0.05cm}+\hspace{-0.05cm}\bm
E\hspace{-0.05cm}=\hspace{-0.05cm}\bm0, \label{closed}
\end{eqnarray}
where $\Delta{\bm W}\hspace{-0.1cm}=\hspace{-0.1cm}\bm
W\hspace{-0.05cm}-\hspace{-0.05cm}\hat{\bm W}$

Next, we have the following theorem to state the stability of the closed-loop system.\\
\textbf{Theorem}: {\em The adaptive impedance control scheme
described by (\ref{control}) and (\ref{NNupdate})
ensures the convergence of the impedance
vector to zero, that is, $\bm z_q\rightarrow\bm 0$ as $t\rightarrow\infty$, if the control parameter $k_g$ is
chosen sufficiently large such that the following condition is satisfied
\begin{eqnarray}
&k_g>b_u\label{condition},
\end{eqnarray}
where $b_u$ denotes the upper bound of $\bm E$.
}

\begin{proof}
First, a Lyapunov-like
candidate is defined as:
\begin{eqnarray}
&V=\frac{1}{2}\bm z_q^T\hspace{-0.05cm}\bm M(\bm q)\hspace{-0.00cm}\bm
z_q\hspace{-0.05cm}+\hspace{-0.05cm}\frac{1}{2}\hspace{-0.05cm}\sum\limits_{j=1}^n\hspace{-0.05cm}\Delta\bm
W_{j}\bm L_{j}^{-1} \Delta\bm W_{j}^T.\label{Lya}
\end{eqnarray}

Next, differentiating $V$ with respect to time, we have:
\begin{eqnarray}
&\dot V\hspace{-0.05cm}=\hspace{-0.05cm}{\bm
z}_q^T\hspace{-0.05cm}\bm M(\bm q)\dot{{\bm
z}}_q\hspace{-0.05cm}+\hspace{-0.05cm}\frac{1}{2}{\bm
z}_q^T\hspace{-0.05cm}\dot{\bm M}(\bm q){\bm
z}_q
-\hspace{-0.05cm}\sum\limits_{j=1}^n\hspace{-0.05cm}\dot{\hat{\bm
W}}_{j}\bm L_{j}^{-1} \Delta\bm W_{j}^T. \label{dLya}
\end{eqnarray}

Substituting the closed-loop equation (\ref{closed}), new
impedance vector (\ref{temp2}) and the update law
(\ref{NNupdate}) into (\ref{dLya}) and using Property (iii), we have:
\begin{eqnarray}
&\dot V\hspace{-0.05cm}=\hspace{-0.05cm}-\bm z_q^T\hspace{-0.05cm}\bm
K_z\bm z_q\hspace{-0.05cm}-\hspace{-0.05cm}\bm z_q^T\{k_g\bm{sgn}(\bm
z_q)
+\Delta\hspace{-0.05cm}{\bm W}\bm\theta(\bm
q,\hspace{-0.05cm}\dot{\bm q},\dot{\bm q}_r, \ddot{\bm q}_r)\hspace{-0.1cm}+\hspace{-0.1cm}\bm
E\}\nonumber\\
&-\sum\limits_{j=1}^n\hspace{-0.05cm}\dot{\hat{\bm
W}}_{j}\bm L_{j}^{-1} \Delta\bm
W_{j}^T
=-\bm z_q^T\hspace{-0.05cm}\bm K_z\hspace{-0.0cm}\bm
z_q\hspace{-0.1cm}-\hspace{-0.05cm}\bm z_q^T\hspace{-0.1cm}\bm
E\hspace{-0.05cm}-\hspace{-0.05cm}k_g\bm
z_q^T\hspace{-0.1cm}\bm{sgn}(\bm z_q
).
\label{dLya2}
\end{eqnarray}
Note that $-\hspace{-0.05cm}\bm z_q^T\hspace{-0.1cm}\bm
E\hspace{-0.05cm}-\hspace{-0.05cm}k_g\bm
z_q^T\hspace{-0.1cm}\bm{sgn}(\bm z_q
)\hspace{-0.1cm}\leq\hspace{-0.1cm}-(k_g\hspace{-0.05cm}-\hspace{-0.05cm}b_u)||\bm
z_q||$. Therefore, if
the condition (\ref{condition}) is satisfied, we have
\begin{eqnarray}
&\dot V\hspace{-0.1cm}\leq\hspace{-0.1cm}-\bm z_q^T\hspace{-0.05cm}\bm
K_z\hspace{-0.0cm}\bm
z_q\hspace{-0.05cm}\leq\hspace{-0.05cm}0.\label{dLya3}
\end{eqnarray}

Hence, we have
$V\hspace{-0.1cm}>\hspace{-0.1cm}0$ and $\dot
V\hspace{-0.1cm}\leq\hspace{-0.1cm}0$. Therefore, $V$ is bounded.
The boundedness of $V$ ensures the boundedness of $\bm z_q$ and
$\Delta\bm\theta$. The
boundedness of $\bm z_q$ ensures the
boundedness of $\bm z$ (see (\ref{temp4})). The boundedness of $\bm z$ ensures the
boundedness of $\Delta\dot{\bm x}$ and $\Delta\bm x$ from
(\ref{vectorz}). From (\ref{temp5}), since $\dot{\bm x}_d$, $\Delta\bm x$, and $\Delta\bm f_l$ are
bounded, $\dot{\bm q}_r$ is bounded. Since both $\bm z_q$ and $\dot{\bm q}_r$ are bounded, $\dot{\bm q}=\bm z_q+\dot{\bm q}_r$ is bounded. Differentiating (\ref{temp5}) with respect to time, it is also concluded that $\ddot{\bm q}_r$ is bounded.
Since $\ddot{\bm q}_r$, $\dot{\bm q}_r$, $\Delta\bm W$,
and $\bm z_q$ are all bounded, $\dot{\bm z}_q$ is
bounded from (\ref{closed}). Therefore, $\bm z_q$ is
uniformly continuous. From (\ref{dLya3}), $\bm z_q$ is bounded in $L^2$.
Then it follows [\ref{slotine}] that $\bm z_q\hspace{-0.1cm}\rightarrow\hspace{-0.1cm}\bm 0$. Hence, the desired impedance model is achieved.
% \cite{aribook,slotine}
\end{proof}

\section{EXPERIMENT}

\begin{figure}[!h]
	\centering\vspace{-0.4cm}
	\begin{minipage}[c]{0.5\textwidth}
		\centering
		\includegraphics[height=5cm,width=8.3cm]{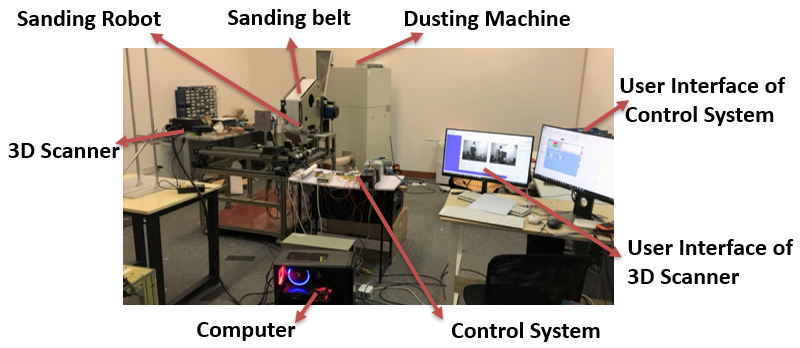}
	\end{minipage}%
	\caption{Overall structure of experient setup}
	\label{Experient}
\end{figure}

\begin{figure}[!h]
    \centering\vspace{0.2cm}
    \begin{minipage}[c]{0.5\textwidth}
        \centering
        \includegraphics[height=3.5cm,width=7.5cm]{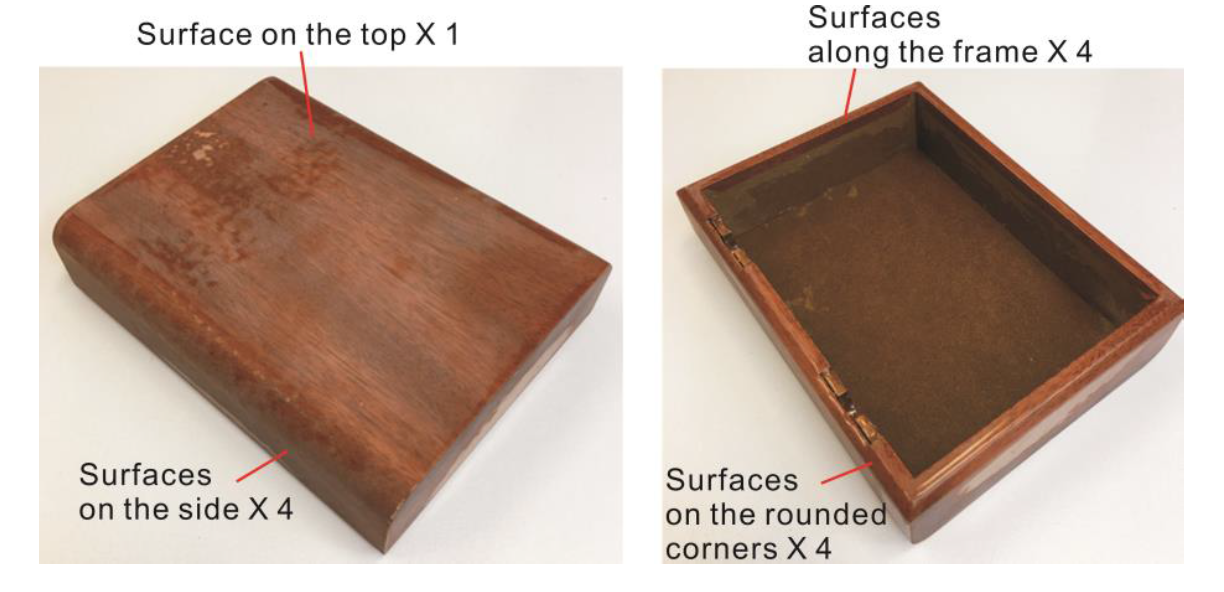}
    \end{minipage}%
    \caption{Sanded object with 13 faces }
    \label{Object}\vspace{-0.7cm}
\end{figure}

The experiment equipment has been set up in The Chinese University of Hong Kong. The whole equipment is shown in Fig. \ref{Experient}. All the joints are driven by AC servo motors which are controlled by IMC30G-E-032PCI control board.  In this experiment, the sanded object is a thirteen faces wooden box shown in Fig.  \ref{Object}. Control system is used to control the four translational and rotational joints  noted as
$ x$ , $ y$ , $ {\theta _1}$ , $ {\theta _2} $ (shown in Fig. \ref{CADmodel}(b)) and the sanding belt.  The visual detecting system includes a 3D scanner, computer with 1080Ti and user interface to collect the original point clouds of the wooden box before and after sanding.
%\subsection{The Scenery of Experiment}
%The sanded object here is a thirteen faces wooden box used for watch case shown in Fig. 7. This experiment will sand out the reflective layer of all faces autonomously and test its quality with structured light camera.
\vspace{-0.2cm}

\begin{figure}[!h]
	\centering
	\vspace{-0.3cm}
	\begin{minipage}[c]{0.5\textwidth}
		\centering
		\includegraphics[height=3.5cm,width=8cm]{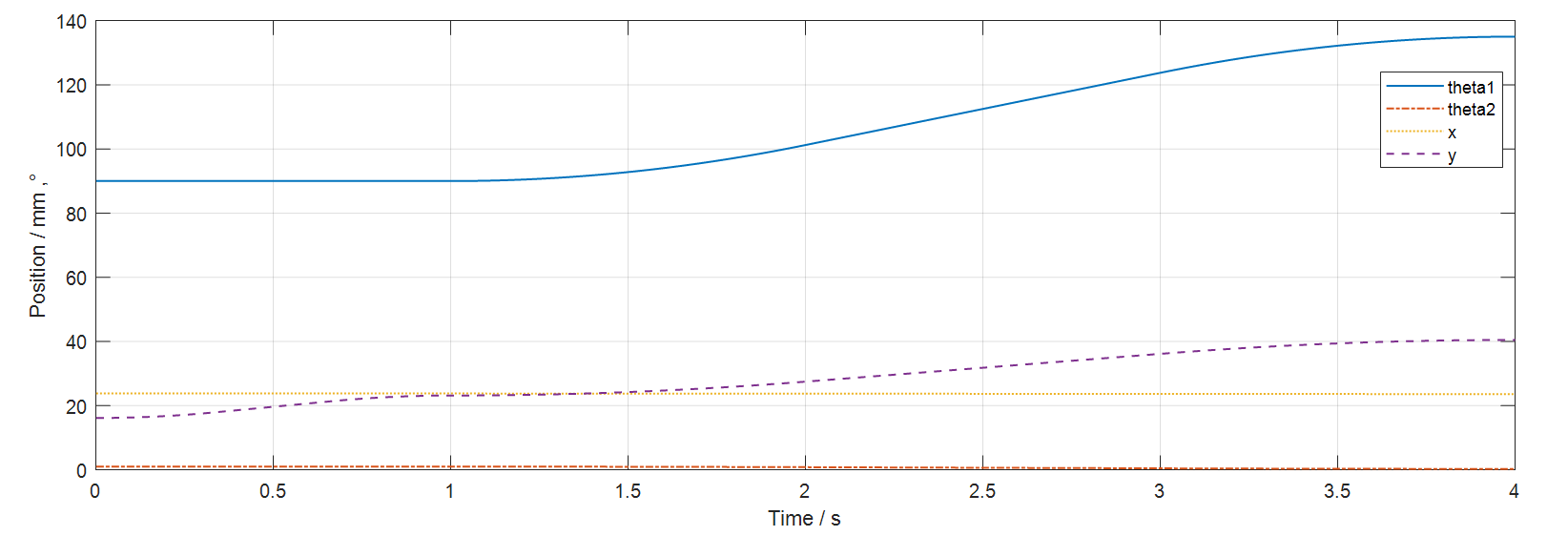}
	\end{minipage}%
	\caption{Single-Query method }
	\label{single}\vspace{-0.3cm}
\end{figure}

Before the experiment, adjust the back plate adjustment tool to parallel compliance with the surface on the top of the box and tune the frequency conversion device to make the sanding belt moving under a proper speed.  When the experiment begins, firstly, the wooden box is scanned by the structured-light camera and then the computer generates the accurate structure of the sanded object after processing. Then the autonomous sanding goes to the motion planning procedure, take one task in single-query method for example, the goal is to change the target surface from one surface on the side to its neighboring surface on the rounded corner. Directly connect initial and goal configurations returns collided path, so the path is updated, and then the desired speed and acceleration are given by LSPB. The results of trajectories of four joints are shown in Fig.\ref{single}.

\begin{figure}[!h]\vspace{-0.2cm}
	\centering\vspace{-0.2cm}
	\begin{minipage}[c]{0.5\textwidth}
		\centering
		\includegraphics[height=3.5cm,width=8cm]{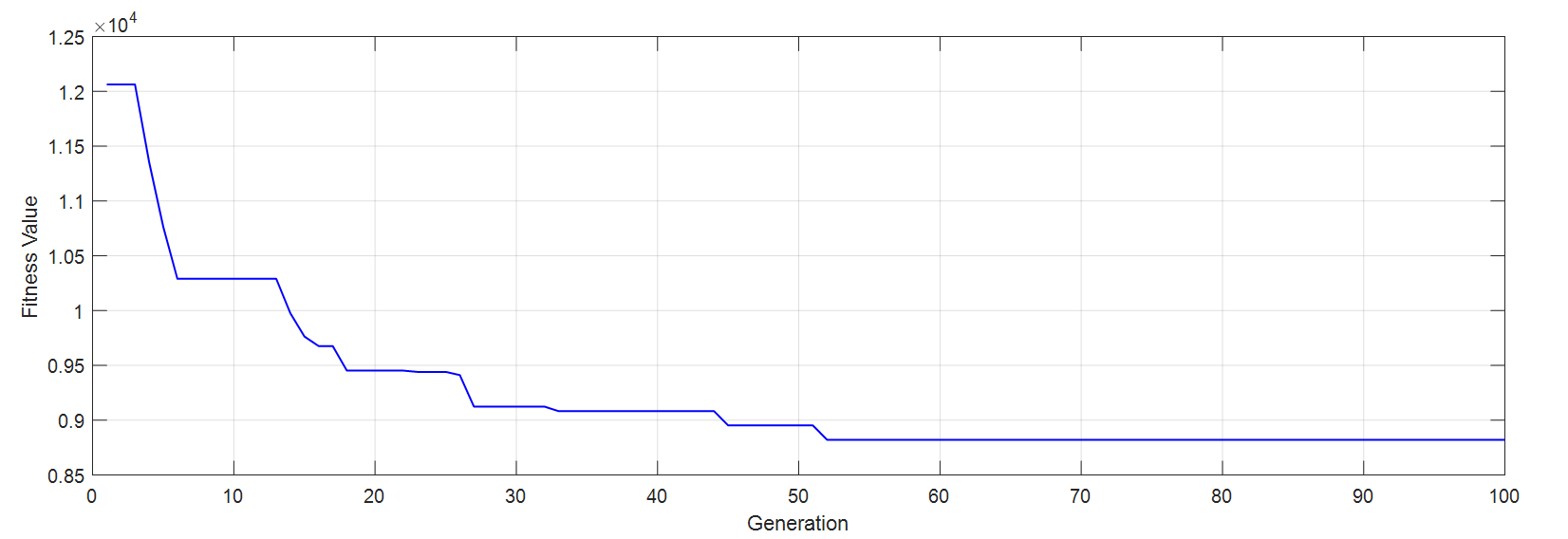} 
	\end{minipage}
	\caption{Sanding Sequence Optimization }
	\label{p4}\vspace{-0.3cm}
\end{figure}

The motion planner uses Genetic Algorithm(GA) to optimize the task sequence in the multi-query method. Set population size as 200, crossover probability as 0.9, mutation probability as 0.1, maximum generation as 100, the optimization result is shown in Fig.\ref{p4}.
%According to the CAD model of the wooden box, with the single-query and multi-query motion planning algorithm, the collision free and data of four joints in different steps will be given out such as shown in table a.  Based on the data in table a, the four joints will move automatically according to this sanding path.

 When the sanded object contact with the sanding belt, the back plate will pop out to make sure the tight contact and then sanding the object with adaptive impedance controller in the sanding direction, the desired contact force during sanding $ f_d $ is -25N, desired position $ x_d $ is 51.5mm. Control parameters $ \Lambda $ = 11.5,
 $ \Gamma $ = 1,  $   K_z$ = 10 , $  k_g$ = 20 and according to (6-8) the desired impedance model was set as $  M_d$ =1, $ C_d$ = 12.5, $  K_d$= 11.5, The impedance error $ z_q $ and control input torque $ u $ are shown in Fig.  \ref{imp}. %the amount of feed is a piecewise linear function.

\begin{figure}[!h]
    \centering\vspace{-0.3cm}
    \begin{minipage}[c]{0.5\textwidth}
        \centering
        \includegraphics[height=3cm,width=7cm]{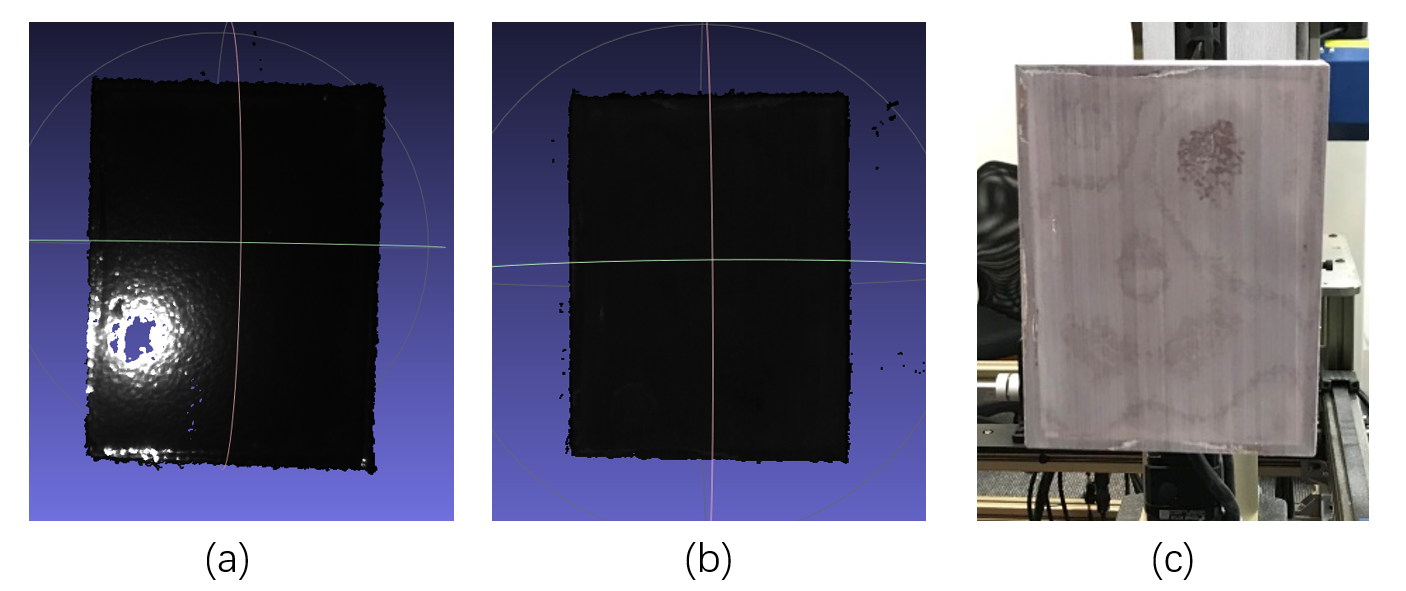}
    \end{minipage}%
    \caption{ (a)  Point cloud data before sanding procedure (b) Point cloud data after sanding procedure (c) The box after sanding}
    \label{CAD}\vspace{-0.3cm}
\end{figure}

\begin{figure}[!h]
	\centering\vspace{-0.3cm}
	\begin{minipage}[c]{0.5\textwidth}
		\centering
		\includegraphics[height=7.5cm,width=8cm]{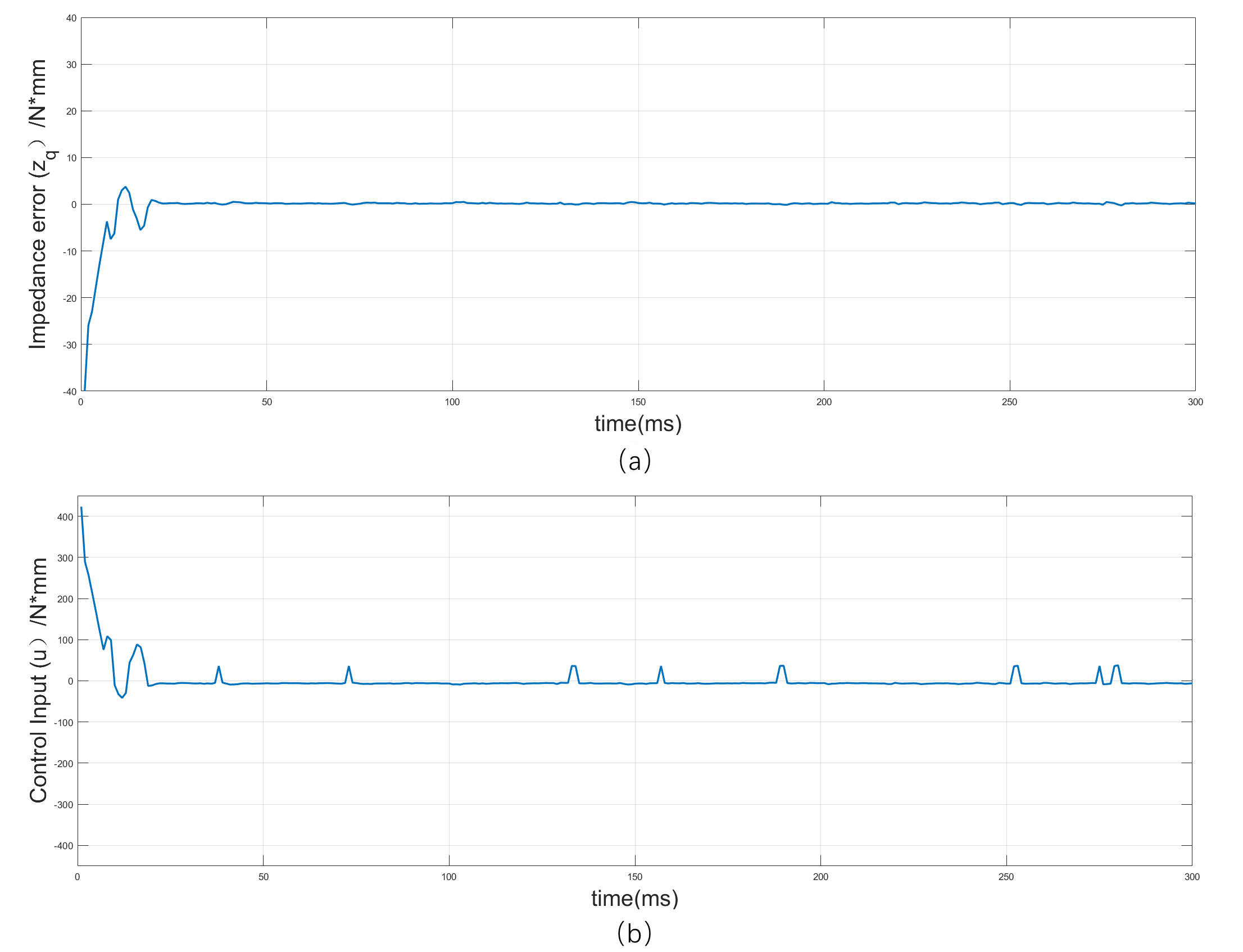}
	\end{minipage}
	\caption{ (a) Impedance error (b) Control input}
	\label{imp}\vspace{-0.3cm}
\end{figure}

 The box after sanding goes back to the vision detection position to evaluate the quality of sanding. The point clouds before and after sanding could be seen in Fig. \ref{CAD}(a) and Fig. \ref{CAD}(b). We can see that the difference in terms of the number of overexposure points between these two point clouds.

%\begin{equation}Feed = \left\{ \begin{array}{l}
%0.5mm,t = 15s\\
%1mm,t = 30s\\
%1.2mm,t = 40s
%\end{array} \right.\end{equation}
In the manufacturing of the wooden box, because of the uneven force and vibration produced during the procedure, it is hard to even skilled workers to sand out the last 0.2mm on the box. However, for this autonomous sanding robot with the structured-light scanner, it is easy to control the force and position of the box without the accurate dynamic model given beforehand. The result of the box after sanding show in Fig. \ref{CAD}(c).

\section{CONCLUSION}
In this paper, a new sanding robot has been developed to automate the labor-intensive works.
The structured-light technology has been employed to scan the unknown object and hence obtain its CAD model online. The model-based planning method has been presented to optimize the motion of the robot end effector. The adaptive impedance controller has been proposed to drive the robot to carry out the sanding task. The stability of the closed-loop system is rigorously proved using Lyapunov methods, with the consideration of uncertain dynamic parameters.
The developed robot has the advantage of highly autonomous capability, in the sense that no human assistance or involvement is required in any steps of the whole procedure of sanding. The performance of the robot has been validated in the sanding experiments on a batch of wooden boxes.

\addtolength{\textheight}{0cm}   % This command serves to balance the column lengths

%%%%%%%%%%%%%%%%%%%%%%%%%%%%%%%%%%%%%%%%%%%%%%%%%%%%%%%%%%%%%%%%%%%%%%%%%%%%%%%%

%%%%%%%%%%%%%%%%%%%%%%%%%%%%%%%%%%%%%%%%%%%%%%%%%%%%%%%%%%%%%%%%%%%%%%%%%%%%%%%%

%%%%%%%%%%%%%%%%%%%%%%%%%%%%%%%%%%%%%%%%%%%%%%%%%%%%%%%%%%%%%%%%%%%%%%%%%%%%%%%%

\end{document}